\documentclass[twocolumn]{article}
\usepackage[utf8]{inputenc}
\usepackage[T1]{fontenc}
\usepackage[english]{babel}
\usepackage{ifpdf,amsmath,amsthm,amssymb,amsfonts,newtxtext,newtxmath} 
\usepackage{array,graphicx,dcolumn,multirow,hevea,abstract,hanging}
\usepackage[labelfont=sc,textfont=sf]{caption}
\usepackage[hyperfootnotes=false,breaklinks=true]{hyperref} 
\usepackage{apacite}
\usepackage[titletoc,toc,title]{appendix}
\usepackage{xcolor}
\usepackage{todonotes}
\usepackage[hyphenbreaks]{breakurl}
\usepackage{booktabs} 
\newcolumntype{L}[1]{>{\raggedright\arraybackslash }p{#1}} 
\newcolumntype{C}[1]{>{\centering\arraybackslash }p{#1}}
\newcolumntype{R}[1]{>{\raggedleft\arraybackslash }p{#1}}
\newcolumntype{d}[1]{D{.}{.}{#1}} 

\let\tempone\itemize
\let\temptwo\enditemize
\let\tempthree\enumerate
\let\tempfour\endenumerate
\renewenvironment{itemize}{\tempone\setlength{\itemsep}{0pt}}{\temptwo}
\renewenvironment{enumerate}{\tempthree\setlength{\itemsep}{0pt}}{\tempfour}

\title{The power of moral words: Loaded language generates framing effects in the extreme dictator game}

\author{
Valerio Capraro\thanks{Middlesex University London. Email: V.Capraro@mdx.ac.uk.}
\and 
Andrea Vanzo\thanks{Heriot-Watt University Edinburgh. Email: a.vanzo@hw.ac.uk}
}

\date{} 
\begin{document} 

\newcommand{\jref}{http://journal.sjdm.org/vol14.1.html}
\newcommand{\jhead}{Judgment and Decision Making, Vol. X, No. Y}
\newcommand{\jdate}{Month Year}
\pagestyle{myheadings} \markright{\protect\small \href{\jref}{\jhead}, \jdate \hfill The power of moral words \qquad}
\begin{htmlonly}
\href{\jref}{\jhead}, \jdate, pp.\
\end{htmlonly}
\twocolumn[
\vspace{-.3in}
{\small \href{\jref}{\jhead}, \jdate, pp.\ WW--ZZ}

\maketitle

\vspace{-3mm}
\begin{onecolabstract}
Understanding whether preferences are sensitive to the frame has been a major topic of debate in the last decades. For example, several works have explored whether the dictator game in the give frame gives rise to a different rate of pro-sociality than the same game in the take frame, leading to mixed results. Here we contribute to this debate with two experiments. In Study 1 ($N=567$) we implement an extreme dictator game in which the dictator either gets \$0.50 and the recipient gets nothing, or the opposite (i.e., the recipient gets \$0.50 and the dictator gets nothing). We experimentally manipulate the words describing the available actions using six terms, from very negative (e.g., stealing) to very positive (e.g., donating) connotations. We find that the rate of pro-sociality is affected by the words used to describe the available actions. In Study 2 ($N=221$) we ask brand new participants to rate each of the words used in Study 1 from ``extremely wrong'' to ``extremely right'' . We find that these moral judgments explain the framing effect in Study 1. In sum, our studies provide evidence that framing effects in an extreme Dictator game can be generated using morally loaded language.

\smallskip
\noindent
{Keywords: framing effect, moral preferences, dictator game, moral judgment}
\end{onecolabstract}\bigskip
]

\saythanks

\setlength{\baselineskip}{12pt plus.2pt}

\section{Introduction}

Understanding whether preferences are sensitive to non-economic cues, such as the name that the experimenter gives to the game or the words used to describe the available actions, has been a major topic of debate in the last two decades.

Earlier works found that people cooperate more in the Prisoner's dilemma when it is called Community Game than when it is called Wall Street Game \cite{kay2003perceptual,liberman2004name}. Yet, it was soon noticed that the presence of social framing effects in the Prisoner's dilemma does not imply that preferences are sensitive to context. It might be that social framing works as a coordination devise, by affecting participants' beliefs about the behavior of the other people involved in the interaction \cite{fehr2006economics}. Consistent with this view, Ellingsen et al. \citeyear{ellingsen2012social} found that social framing effects in the Prisoner's dilemma vanish when the game is played sequentially, suggesting that social cues primarily works by changing participants' beliefs.

To shed light on this issue, in more recent years scholars have started investigating whether social framing can affect people's choices not in strategic games, but in social decision problems. In this context, the decision maker can affect the outcome of one or more other people, but these people have no way to affect the decision maker's outcome. Since in these situations beliefs play no role, any framing effect would be uniquely driven by a chance in preferences. 

A stream of research has focused on whether the words used to describe the available actions can affect decision makers' choices. The empirical evidence has shown that indeed they can, at least in some contexts. Eriksson et al. \citeyear{eriksson2017costly} have found that rejection rates in the Ultimatum game depend on whether the rejection option is called ``payoff reduction'' or ``costly punishment''. In the former case, responders are less likely to reject the proposer's offer, and this effect appears to be driven by moral concerns. Similarly, Capraro and Rand \citeyear{capraro2018right} and Tappin and Capraro \citeyear{tappin2018doing} have found that behavior in the Trade-Off game, i.e. a decision problem in which decision-makers have to decide between an equitable and an efficient allocation of resources, highly depends on which choice is presented as being the ``right thing to do''.

Particularly interesting is the situation regarding the Dictator game (DG). The DG, whereby the \textit{dictator} has to decide how to unilaterally divide a sum of money between her/himself and the \textit{recipient}, is considered to be the standard measure of altruistic behavior and, for this reason, is among the most studied decision problems in social science \cite{forsythe1994fairness,eckel1996altruism,engel2011dictator,rand2016social}. Studies exploring framing effects in the DG have led to mixed results. Previous works have mainly focused on whether the DG in the \textit{Take} frame gives rise to higher pro-sociality than the same game in the \textit{Give} frame. Swope et al. \citeyear{swope2008social} have found that participants tend to be more pro-social in the DG in the \textit{Take} frame than in the DG in the \textit{Give} frame. In the \textit{Take} frame, the endowment was initially given to the recipient, and dictators could take any amount. Krupka and Weber \citeyear{krupka2013identifying} have replicated this framing effect in a context in which the initial endowment is split equally among the dictator and the recipient. Additionally, Krupka and Weber \citeyear{krupka2013identifying} have shown that this framing effect is driven by a change in the perception of what is the ``socially appropriate thing to do''. However, the framing effect when passing from the \textit{Take} frame to the \textit{Give} frame has not been replicated by several other works \cite{dreber2013people,grossman2015giving,halvorsen2015dictators,hauge2016keeping,goerg2017framing}, casting doubts on the very existence of framing effects in the DG.\footnote{Related to this literature, some studies have shown that extending the choice set of the standard DG to include a taking option has the effect of decreasing giving \cite{list2007interpretation,bardsley2008dictator,cappelen2013give,zhang2013interpretation}. Eichenberger and Oberholzer-Gee \citeyear{eichenberger1998rational} have found that the amount \emph{given} by students who perform well in a test to students who perform bad in the same test is greater than the amount \emph{not taken} by students who perform bad in a test from students who perform well. Chang, Chen and Krupka \citeyear{chang2017rhetoric} have found that the DG in the tax frame gives rise to more pro-sociality than the standard DG.}

Here we focus on the existence of framing effects in the DG with two experiments. In Study 1, we implement an extreme DG, where the dictator can either allocate \$0.50 to himself and \$0 to the recipient, or the other way around, s/he can allocate \$0.50 to the recipient and \$0 to himself. We then manipulate the words used to describe the two available actions through six different frames. We find that this word manipulation significantly impacts participants' decisions. In Study 2 we ask brand new participants to rate all the words used to describe the available actions in Study 1, from ``extremely wrong'' to ``extremely right''. We show that the rate of pro-sociality in a given frame of Study 1 can be predicted by the difference between the rating of the word associated, in the same frame, to the pro-social action and the rating of the word associated, in the same frame, to the pro-self action.

\section{Study 1}

We consider a variant of the standard DG. In the standard DG, the dictator has to allocate an amount of money (e.g., \$0.50) between her/himself and another participant (typically anonymous). The recipient has no choice and only gets what the dictator decides to give. Conversely, in our \textit{extreme} version, the dictator can opt only for the two extreme options: either s/he gets the whole \$0.50 (and the recipient gets nothing) or the other way around, i.e., the recipient gets the \$0.50 and the dictator gets nothing. The reason why we choose this extreme variant, rather than the standard one, is to be able to write the instructions of the decision problem in several different treatments by changing just one word. This would have been hard, if not impossible, with the classical ``continuous'' variant, as it will be clear later. In this extreme DG, we hypothesize that the words used to describe the available actions will impact participants' choices, beyond the economic consequences that these actions bring about. The intuition is that, under the same economic conditions, naming the self-regarding action through an extremely ``negative'' expression as \textit{Steal from the other participant}, rather than using the more ``neutral'' expression \textit{Take from the other participant}, will significantly impact the final decision. 

\subsection{Experimental Design}
Participants are randomly assigned to one of twelve conditions. In the \textit{Steal vs Don't steal} condition, they are told that there are \$0.50 available and they have to choose between two possible actions: \textit{Steal from the other participant}, so that they would get the \$0.50 and the other participant would get nothing; or \textit{Don't steal from the other participant}, so that the other participant would get the \$0.50, while they would get nothing. It is made explicit that the other participant has no choice and will be really paid according to the decision made. Participants are also asked two comprehension questions, to make sure they understand the decision problem. One question asks which choice would maximize their payoff; the other one asks which choice would maximize the other participant's payoff. Participants failing either or both comprehension questions are automatically excluded from the survey. Those who pass the comprehension questions are asked to make the real choice. The \textit{Don't steal vs Steal} condition is similar: the only difference is that we switched the order of presentation of the options, in order to account for order effects. The conditions \textit{Take vs Don't take}, \textit{Don't take vs Take}, \textit{Demand vs Don't demand}, \textit{Don't demand vs Demand}, \textit{Give vs Don't give}, \textit{Don't give vs Give}, \textit{Donate vs Don't donate}, \textit{Don't donate vs Donate}, \textit{Boost vs Don't boost}, \textit{Don't boost vs Boost} are analogous.
Hence, the instructions of one frame differ from those of another frame only in one word (see Appendix for full instructions). This justifies our choice of an extreme variant of the Dictator game, rather than the classical ``continuous'' variant. A moment of reflection shows that it is very hard, if not impossible, to write, for example, the \textit{Take} frame and the \textit{Give} frame in a ``continuous'' form using instructions differing only in one word.\footnote{The choice of the words ``steal'', ``take'', ``demand'', ``donate'', ``give'', and ``boost'' was motivated by Sentiment Analysis \cite{pang2004sentimental,pang2002thumbs,liu2007low,mcglohon2010star,vanzo2014context}. In particular, we based our choice on a lexical resource known as SentiWordNet \cite{baccianella2010sentiwordnet,esuli2006sentiwordnet}, which divides the English vocabulary in synsets (i.e., sets of synonyms). We selected six synsets containing words that are particularly suitable to describe actions in an extreme dictator game: \textit{\textbf{steal}\#1} = ``take without the owner's consent''; \textit{\textbf{take}\#8} = ``take into one's possession''; \textit{\textbf{demand}\#1} = ``request urgently and forcefully''; \textit{\textbf{give}\#3} = ``transfer possession of something concrete or abstract to somebody''; \textit{\textbf{donate}\#1} = ``give to a charity or good cause''; and \textit{\textbf{boost}\#2} = ``be beneficial to''.}

A standard demographic questionnaire concludes the survey. At the end of the questionnaire, participants are communicated the completion code with which they can submit the survey to Amazon Mechanical Turk (AMT) and claim their payment. Payoffs are computed and paid on top of their participation fee (\$0.50). The other participant was selected at random from the same sample. Therefore, participants received a payment both from their choice as decision makers, and from being in the role of the ``other participant'' for a different participant. It is worth noticing that when making their choice, participants were not informed that they would also be in the role of the ``other participant'' to avoid having this affect their decision. We refer to the Appendix for verbatim experimental instructions.

\subsection{Results}
\subsubsection{Participants}
We recruited US based participants on the online platform Amazon Mechanical Turk (AMT).\footnote{Several works have shown that data gathered using AMT are of no less quality than data collected using the standard laboratory \cite{arechar2018conducting,branas2018gender,horton2011online,paolacci2014inside,paolacci10runningexperiments,rand2012promise}.} We collected $727$ observations and cleaned this dataset through the following two operations. First, we looked for multiple observations by checking multiple IP addresses and multiple Turk IDs. In case we found multiple observations, we kept only the first one as determined by the starting date and we deleted all the remaining ones. Second, we discarded all participants who failed either or both comprehension questions. After these operations, we remained with a sample of $N=567$ participants ($52.7\%$ females, mean age = $37.6$ years). Thus, in total, $22\%$ of the participants have been eliminated from the analysis. This is in line with previously published studies using AMT \cite{horton2011online}.

\subsubsection{Pro-Sociality}

To analyze this sample, we first build a binary variable, named \textit{Pro-Sociality}, which takes value $1$ whenever a participant allocates the \$0.50 to the other person, and $0$ whenever a participant allocates the \$0.50 to him/herself. Average \textit{Pro-Sociality} across the whole sample is $0.141$, that is, $14.1\%$ of the participants act pro-socially. \textit{Pro-Sociality} within a frame turns out to not depend on the order in which the options are presented, e.g., average \textit{Pro-Sociality} in the condition \textit{Steal vs Don't steal} is not statistically different from average \textit{Pro-Sociality} in the condition \textit{Don't steal vs Steal}. Similarly for all other frames. Thus, in what follows we collapse across order and we name $w$ the union between condition \textit{w vs Don't w} and \textit{Don't w vs w}.

\subsubsection{Framing effect}
\begin{figure}
\centering
\includegraphics[width=1\linewidth]{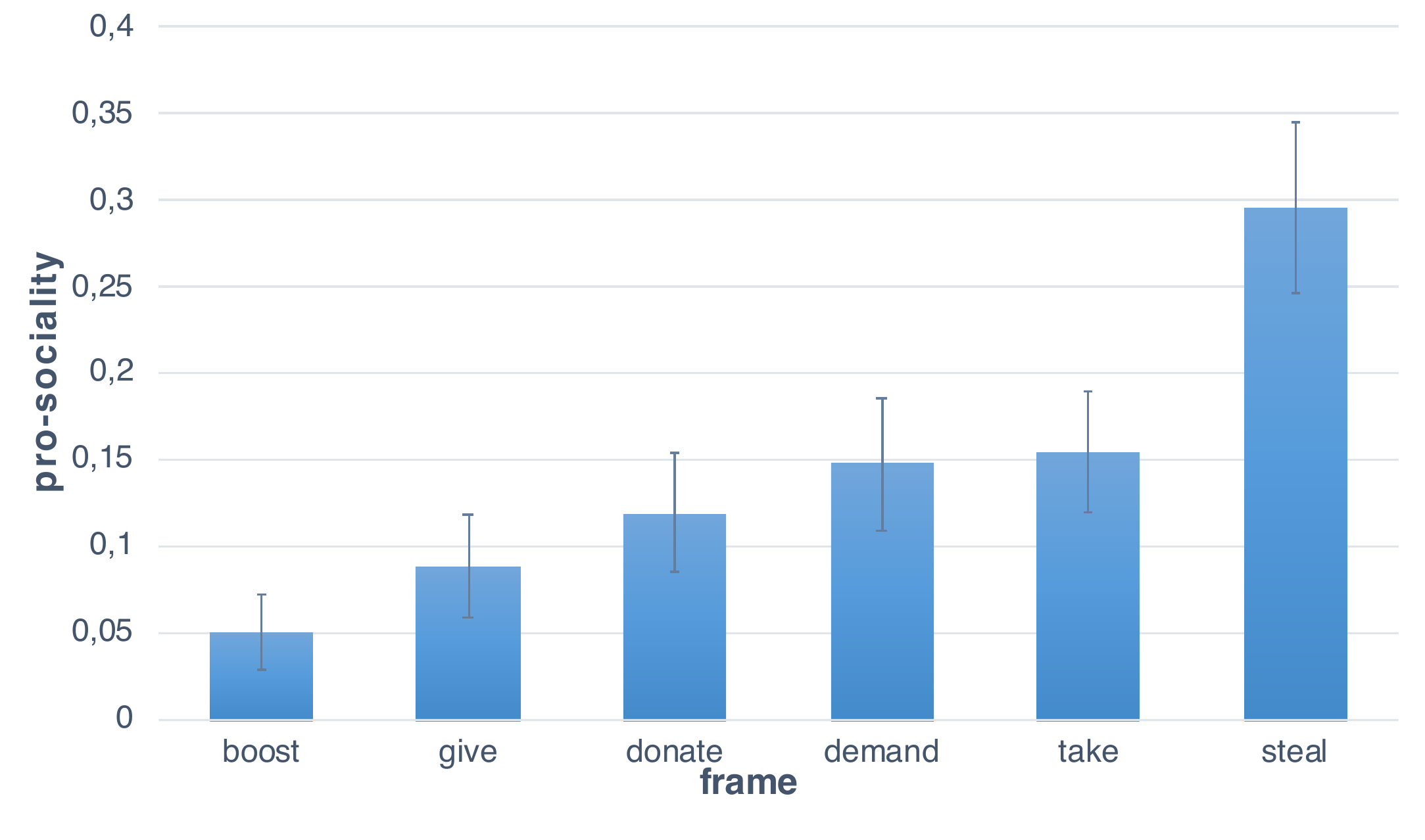}
\caption{\textit{Pro-Sociality} across frames in Study 1. Error bars represent standard errors of the means.}
\label{fig:figure1}
\end{figure}

We now test our hypothesis that the words used to describe the available actions affect \textit{Pro-Sociality}. Figure \ref{fig:figure1} provides visual evidence that there is significant variability across conditions: the minimum of the variable \textit{Pro-Sociality} is attained in the \textit{Boost} frame ($5.0\%$), while its maximum is attained in the \textit{Steal} frame ($29.5\%$). Table \ref{tab:regression} reports coefficients and standard errors of pairwise logit regressions. In short, it emerges that the \textit{Steal} frame produces an amount of pro-sociality that is statistically significantly higher than that of any other frames. Conversely, the \textit{Boost} frame gives rise to an amount of pro-sociality that is numerically lower than that of any other frames; however, the difference is statistically significant only versus the \textit{Demand}, \textit{Take}, and \textit{Steal} frames, while it is marginally significant versus the \textit{Donate} frame.
The other frames numerically lie between the \textit{Boost} frame and the \textit{Steal} frame, but pairwise differences are not statistically significant.

\begin{table}
\centering
\caption{Pairwise logit regression predicting the effect on \textit{Pro-Sociality} of passing from one frame of Study 1 to the other one. We report coefficients and, in brackets, standard errors. Significance thresholds: *: p < 0.1, **: p < 0.05, ***: p < 0.01.\label{tab:regression}}
\begin{tabular}{llllll}
	\toprule
	& \textbf{Give} & \textbf{Donate} & \textbf{Demand} & \textbf{Take} & \textbf{Steal}\\
	\midrule
	\multirow{2}{*}{\textbf{Boost}}  & -0.607 & -0.937* & -1.181** & -1.234** & -2.065***\\
	& (0.590) & (0.560) & (0.548) & (0.529) & (0.515)\\
	\multirow{2}{*}{\textbf{Give}} &  & -0.331 & -0.575 & -0.628 & -1.458***\\
	&  & (0.490) & (0.477) & (0.455) & (0.438)\\
	\multirow{2}{*}{\textbf{Donate}} &  &  & -0.244 & -0.297 & -1.128***\\
	&  &  & (0.440) & (0.416) & (0.397)\\
	\multirow{2}{*}{\textbf{Demand}} &  &  &  & -0.053 & -0.884**\\
	&  &  &  & (0.400) & (0.381)\\
	\multirow{2}{*}{\textbf{Take}} &  &  &  &  & -0.830**\\
	&  &  &  &  & (0.352)\\
	\bottomrule
\end{tabular}
\end{table}




\section{Study 2}

Study 1 shows that the words used to describe the available actions in an extreme dictator game can significantly impact participants' decisions. How can we explain this result? Since different frames have the same economic structure, this finding is inconsistent with standard theories of social preferences, which assume that people's utility function depends only on the economic consequences of the available actions \cite{fehr1999theory,bolton2000erc,charness2002understanding,engelmann2004inequality}. One potential explanation is that participants have moral preferences: they compare the moral goodness of the pro-social action to the moral badness of the pro-self action, and then act pro-socially only when the resulting difference is large enough to counterbalance the cost of the pro-social action. This explanation makes the empirically testable prediction that there is a correlation between pro-sociality in frame $w$ and the difference between the moral judgment associated to the pro-social action in frame $w$ and the pro-self action in frame $w$. In Study 2 we test this hypothesis.

\subsection{Experimental design}

Participants are presented, in random order, the instructions of all the six frames of the extreme Dictator game in Study 1. After reading the instructions of the extreme Dictator game, for each frame word $w\in\{Steal, Demand, Take, Give, Boost, Donate\}$, e.g. $w = Steal$, participants are asked the following two questions in random order:

\begin{itemize}
	\item \emph{From a moral point of view, how would you judge the choice: to steal?}
	\item \emph{From a moral point of view, how would you judge the choice: not to steal?}
\end{itemize}

Answers are collected using a 5-point Likert scale with: 1 = ``extremely wrong'', 2 = ``somewhat wrong'', 3 = ``neutral'', 4 = ``somewhat right'', and 5 = ``extremely right''. After answering all these twelve (two for each frame) questions, participants enter the demographic questionnaire ending the survey. Note that, in this study, participants do not make any decision.\footnote{We deliberately decided to collect moral judgments using a different sample to avoid that any correlation be driven by subjects justifying their choice in terms of morality.}






\subsection{Results}

\subsubsection{Participants}
We recruited 250 US based participants on AMT. None of them participated in Study 1. After eliminating multiple IP addresses and multiple TurkIDs, we remained with 221 observations (female = 41.9\%, mean age = 34.25).

\subsubsection{Predicting pro-sociality from self-reported moral judgments}\label{suse:predict_judgm}

We aim at showing that the rate of pro-sociality in Study 1 in frame $w$ can be predicted by the difference between the moral judgment of the pro-social action in frame $w$ and the moral judgment of the pro-self action in frame $w$.

To this end, we define the following sets. Let $W^+=\{\textit{Don't steal}, \textit{Don't take}, \textit{Don't demand}, \textit{Boost}, \textit{Give}, \textit{Donate} \}$ denote the set of wordings corresponding to the pro-social action. Similarly, let us also define the set $W^-=\{\textit{Steal}, \textit{Take}, \textit{Demand}, \textit{Don't boost}, \textit{Don't give}, \textit{Don't donate} \}$, to be the set of wordings corresponding to the pro-self action. For each $w\in W^+$, let $\lnot w$ be the corresponding word in $W^-$, that is, for example, $\lnot(\textit{Don't steal}) = \textit{Steal}$. 

For each $w\in W^+$, we define the polarization of $w$ to be $\textit{Pol(w)}=\textit{Judgm(w)}-\textit{Judgm}(\lnot w)$, where $Judgm(w)$ is the self-reported moral judgment of $w$, collected in Study 2 and normalized between -1 and 1, with 0 corresponding to ``neutral''.

\begin{figure}[h]
\centering
\includegraphics[width=1\linewidth]{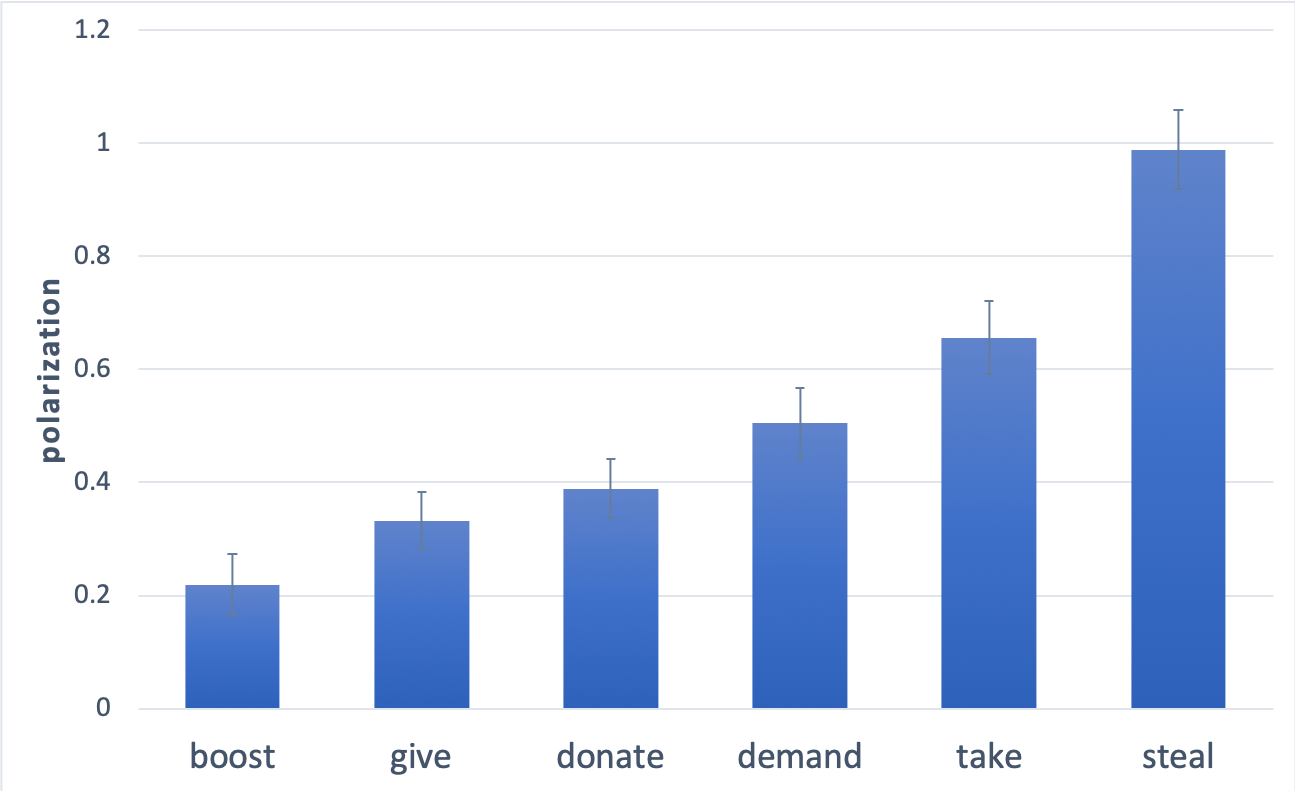}
\caption{\textit{Polarization} across conditions in Study 2. Error bars represent standard errors of the means.}
\label{fig:figure2}
\end{figure}

Figure \ref{fig:figure2} reports all the polarizations. When comparing it to Figure  \ref{fig:figure1}, one immediately notices that the pattern of the polarizations is very similar to the pattern of pro-sociality. To show that the polarizations indeed explain pro-sociality in Study 1, we proceed in two steps. First, we note that logit regression finds an overall significant effect of the variable \textit{Pol(w)} on \textit{Pro-Sociality} (coeff = 2.218, z = 4.867, p $< .001$). However, an overall effect does not automatically imply that the details of the effect are well explained. Therefore, we next explore the effect of \textit{Pol(w)} on \textit{Pro-Sociality} for every pair $w_1,w_2\in W^+$. Table \ref{tab:regression_krupka} reports coefficients and standard errors of pairwise logit regressions predicting \textit{Pro-Sociality} as a function of \textit{Pol(w)} for every pair $w_1,w_2\in W^+$. Comparing Table \ref{tab:regression_krupka} with Table \ref{tab:regression}, we observe that all the significance levels are exactly the same, suggesting that indeed \textit{Pol(w)} is a good predictor of \textit{Pro-Sociality}, even at a point-wise level. To strengthen this conclusion, we observe that the coefficients in Table \ref{tab:regression_krupka} are all similar in magnitude, as shown by meta-analysis of the coefficients, which finds an overall effect size of $2.238$, statistically significant ($z=7.70, p < .001$), and, crucially, no heterogeneity in the coefficients ($p=0.977$). Note that the coefficient obtained via meta-analysis (2.238) is very similar to the coefficient obtained via logit regression (2.218), suggesting that, in our experiment, the same coefficient for the variable \textit{Pol(w)} can explain the rate of pro-sociality in all the frames of Study 1.

\begin{table}
\centering
\caption{Pairwise logit regression predicting the effect of \textit{Pol(w)} on \textit{Pro-Sociality}, for every pair $w_1,w_2\in W^+$. We report coefficients and, in brackets, standard errors. Significance thresholds: *: p < 0.1, **: p < 0.05, ***: p < 0.01.\label{tab:regression_krupka}}
\begin{tabular}{llllll}
	\toprule
	& \textbf{Give} & \textbf{Donate} & \textbf{Demand} & \textbf{Take} & \textbf{Steal}\\
	\midrule
	\multirow{2}{*}{\textbf{Boost}}  & 5.514 & 5.858* & 4.219** & 2.805** & 2.681***\\
	& (5.362) & (3.502) & (1.960) & (1.203) & (0.669)\\
	\multirow{2}{*}{\textbf{Give}} &  & 6.614 & 3.381 & 1.902 & 2.209***\\
	&  & (9.807) & (2.805) & (1.378) & (0.663)\\
	\multirow{2}{*}{\textbf{Donate}} &  &  & 2.033 & 1.061 & 1.848***\\
	&  &  & (3.665) & (1.484) & (0.651)\\
	\multirow{2}{*}{\textbf{Demand}} &  &  &  & 0.332 & 1.803**\\
	&  &  &  & (2.498) & (0.777)\\
	\multirow{2}{*}{\textbf{Take}} &  &  &  &  & 2.516**\\
	&  &  &  &  & (1.068)\\
	\bottomrule
\end{tabular}
\end{table}

\section{Discussion}

In Study 1 we have shown that the words used to describe the available actions in an extreme Dictator game can significantly alter subjects' behavior. Study 2 shows that this framing effect is consistent with a theory of moral preferences, according to which participants compare the moral goodness of the pro-social action to the moral badness of the pro-self action, and then act pro-socially only when their difference is large enough to counterbalance the cost of the pro-social action. In this view, the words in Study 1 affect behavioral outcomes by affecting the moral judgments associated to the available actions.

A number of papers have explored whether the Dictator Game in the \textit{Take} frame gives rise to greater pro-sociality than the same game in the \textit{Give} frame. Swope et al. \citeyear{swope2008social} and Krupka and Weber \citeyear{krupka2013identifying} have found that, indeed, participants tend to be more pro-social in the \textit{Take} frame than in the \textit{Give} frame. Moreover, Krupka and Weber \citeyear{krupka2013identifying} have shown that the rate of pro-sociality can be predicted by what they called ``social appropriateness'' of an action. However, the framing effect when passing from the \textit{Take} frame to the \textit{Give} frame has not been replicated by several other works \cite{dreber2013people,grossman2015giving,halvorsen2015dictators,hauge2016keeping,goerg2017framing}. This mixed results thus left open the question about whether framing effects actually exist in the Dictator game and whether they can be actually explained in terms of ``social appropriateness'' or similar constructs. Our work sheds light on this topic. We have indeed found that the words used to describe the available actions in an extreme Dictator game can alter people's behavior, and that, somewhat in line with Krupka and Weber \citeyear{krupka2013identifying}, that the rate of pro-sociality can be predicted by the moral judgments associated to the available actions. However, in line with \cite{dreber2013people,grossman2015giving,halvorsen2015dictators,hauge2016keeping,goerg2017framing} we have found that the \textit{Take} frame does not give rise to a rate of pro-sociality significantly higher than the \textit{Give} frame.

The fact that participants care about doing what they think is the morally right thing has been shown in several contexts in the last years. For example, empirical evidence has been presented to support the hypotheses that (i) altruistic behavior in the Dictator game and cooperative behavior in the Prisoner's dilemma are partly driven by a desire to do the ``right thing'' \cite{capraro2018right,tappin2018doing}, (ii) differences in behavior in the Dictator game in the ``give'' frame vs the ``take'' frame are driven by a change in people's perception about what is the appropriate thing to do \cite{krupka2013identifying}, (iii) a change in the perception of what is the morally wrong thing to do can generate framing effects among Ultimatum game responders  \cite{eriksson2017costly}, (iv) moral reminders increase pro-social behavior in both the Dictator game and the Prisoner's dilemma \cite{branas2007promoting,capraro2017right,dal2014right}. A number of theoretical models have also been introduced to formalize people's tendency towards doing the right thing \cite{alger2013homo,brekke2003economic,dellavigna2012testing,huck2012social,kessler2012norms,kimbrough2016norms,krupka2013identifying,lazear2012sorting,levitt2007laboratory}. Our work adds to this literature by showing that small changes in the language used to describe the available actions can affect the moral judgments associated to those actions, which ultimately affect participants' decisions.

Related to our work is also the game-theoretical literature on language-based games. In fact, the idea that the language used to describe the available strategies can affect people's behavior has been supported also by game theorists. Bjorndahl, Halpern \& Pass \citeyear{bjorndahl2013language} have argued that outcome-based preferences do not suffice to explain some human interactions, which are instead best understood by defining the utility function on the underlying language used to describe the game. Motivated by this observation, they have defined the class of language-based games and have studied a generalization of Nash equilibrium and rationalizable strategies on these games. 

Of course, our results should be interpreted within the natural limitations of our experiment: an extreme variant of the Dictator Game with \$0.50 at stake. For example, although previous research has found very little evidence of stake effects in several games involving pro-sociality when stakes are not too high \cite{forsythe1994fairness,carpenter2005effect,johansson2005does,branas2018gender,larney2019stake}, other studies have found evidence that pro-sociality decreases at very high stakes \cite{carpenter2005effect,andersen2011stakes}. Along these lines, it might be possible that, when facing very high stakes, participants' behavior becomes less influenced by the words being used to describe the available actions. Understanding the boundary conditions of the effect of the words used to describe the available actions on participants' behavior is a primary direction for future research. 
Moreover, our results are silent about the channel through which words affect moral judgments. One possibility is that the words generate a continuous endowment effect \cite{reb2007possession} which ultimately affects participants' choices. This is certainly a topic that needs to be thoroughly investigated.

In conclusion, our data show that the words used to describe the available actions can affect people's decisions in extreme dictator games. Future research should explore the boundary conditions of this effect and the conceptual channels through which words affect moral judgments.

\bibliographystyle{apacite}
\bibliography{bibliography.bib}

\begin{appendices}

\section{Experimental Instructions of Study 1}
\subsection{\textit{Steal vs Don't steal} condition}
There are 50 cents available. You have to choose between two possible actions:
\begin{itemize}
	\item STEAL FROM THE OTHER PARTICIPANT: In which case, you get the 50 cents and the other participant gets 0 cents;
	\item DON'T STEAL FROM THE OTHER PARTICIPANT: In which case, you get 0 cents and the other participant gets 50 cents.
\end{itemize}
The other participant has no choice and will be paid according to your decision. No deception is used. You and the other participant will be paid according to your decision.

Here are some questions to ascertain that you understand the rules. Remember that you have to answer all of these questions correctly in order to get the completion code. If you fail any of them, the survey will automatically end and you will not get any payment.\footnote{A skip logic in the survey eliminated from the survey automatically all participants providing the wrong answer}
\begin{itemize}
	\item What choices should YOU make in order to maximize YOUR gain?
	[\textit{Available answers: Steal from the other participant -- Don't steal from the other participant.}]
	\item What choice should YOU make in order to maximize the OTHER PARTICIPANT's gain?
	[\textit{Available answers: Steal from the other participant -- Don't steal from the other participant.}]
\end{itemize}
Congratulations, you passed all comprehension questions. It is now time to make your choice.
\begin{itemize}
	\item What is your choice?
	[\textit{Available options: Steal from the other participant (50 cents for you, 0 cents for the other participant) / Don't steal from the other participant (0 cents for you, 50 cents for the other participants)}]
\end{itemize}

\subsection{\textit{Don't steal vs Steal} condition}
\textit{Identical to the previous one, with the only difference that the word ``Steal'' was replaced by ``Don't Steal'', and the words ``Don't steal'' were replaced by ``Steal''. Payoffs were changed accordingly.}\footnote{An identical scheme has been adopted for the \textit{Take vs Don't take}, \textit{Don't take vs take}, \textit{Demand vs Don't demand}, \textit{Don't demand vs Demand}, \textit{Give vs Don't give}, \textit{Don't give vs Give}, \textit{Donate vs Don't donate}, \textit{Don't donate vs Donate}, \textit{Boost vs Don't boost}, and \textit{Don't boost vs Boost} conditions.}

\section{Experimental Instructions of Study 2}

\subsection{\textit{Moral judgment} condition}

[The following questions were asked in random order]

\textit{Steal screen}

Imagine that there are 50 cents available and that you have to choose between two possible actions:

Steal from the other participant: In which case, you get the 50 cents and the other participant gets 0 cents

Don't steal from the other participant: In which case, you get 0 cents and the other participant gets 50 cents.

Having this situation in mind, please answer the following questions [presented in random order]:

From a moral point of view, how would you judge the choice: to steal? [Available answers: Extremely wrong / Somewhat wrong / Neutral / Somewhat right / Extremely right]

From a moral point of view, how would you judge the choice: not to steal? [Available answers: Extremely wrong / Somewhat wrong / Neutral / Somewhat right / Extremely right]

\bigskip

\textit{Take screen}

Imagine that there are 50 cents available and that you have to choose between two possible actions:

Take from the other participant: In which case, you get the 50 cents and the other participant gets 0 cents

Don't take from the other participant: In which case, you get 0 cents and the other participant gets 50 cents.

Having this situation in mind, please answer the following questions [presented in random order]:

From a moral point of view, how would you judge the choice: to take? [Available answers: Extremely wrong / Somewhat wrong / Neutral / Somewhat right / Extremely right]

From a moral point of view, how would you judge the choice: not to take? [Available answers: Extremely wrong / Somewhat wrong / Neutral / Somewhat right / Extremely right]

\bigskip

\textit{Demand screen}

Imagine that there are 50 cents available and that you have to choose between two possible actions:

Demand from the other participant: In which case, you get the 50 cents and the other participant gets 0 cents

Don't demand from the other participant: In which case, you get 0 cents and the other participant gets 50 cents.

Having this situation in mind, please answer the following questions [presented in random order]:

From a moral point of view, how would you judge the choice: to demand? [Available answers: Extremely wrong / Somewhat wrong / Neutral / Somewhat right / Extremely right]

From a moral point of view, how would you judge the choice: not to demand? [Available answers: Extremely wrong / Somewhat wrong / Neutral / Somewhat right / Extremely right]

\bigskip

\textit{Give screen}

Imagine that there are 50 cents available and that you have to choose between two possible actions:

Give from the other participant: In which case, you get the 0 cents and the other participant gets 50 cents

Don't give from the other participant: In which case, you get 50 cents and the other participant gets 0 cents.

Having this situation in mind, please answer the following questions [presented in random order]:

From a moral point of view, how would you judge the choice: to give? [Available answers: Extremely wrong / Somewhat wrong / Neutral / Somewhat right / Extremely right]

From a moral point of view, how would you judge the choice: not to give? [Available answers: Extremely wrong / Somewhat wrong / Neutral / Somewhat right / Extremely right]

\bigskip

\textit{Donate screen}

Imagine that there are 50 cents available and that you have to choose between two possible actions:

Donate from the other participant: In which case, you get the 0 cents and the other participant gets 50 cents

Don't donate from the other participant: In which case, you get 50 cents and the other participant gets 0 cents.

Having this situation in mind, please answer the following questions [presented in random order]:

From a moral point of view, how would you judge the choice: to donate? [Available answers: Extremely wrong / Somewhat wrong / Neutral / Somewhat right / Extremely right]

From a moral point of view, how would you judge the choice: not to donate? [Available answers: Extremely wrong / Somewhat wrong / Neutral / Somewhat right / Extremely right]

\bigskip

\textit{Boost screen}

Imagine that there are 50 cents available and that you have to choose between two possible actions:

Boost the other participant: In which case, you get the 0 cents and the other participant gets 50 cents

Don't boost the other participant: In which case, you get 50 cents and the other participant gets 0 cents.

Having this situation in mind, please answer the following questions [presented in random order]:

From a moral point of view, how would you judge the choice: to boost? [Available answers: Extremely wrong / Somewhat wrong / Neutral / Somewhat right / Extremely right]

From a moral point of view, how would you judge the choice: not to boost? [Available answers: Extremely wrong / Somewhat wrong / Neutral / Somewhat right / Extremely right]




\end{appendices}

\end{document}